\documentclass[aps,prx,twocolumn,nofootinbib,groupedaddress,superscriptaddress,longbibliography]{revtex4-2}
\usepackage{amssymb}
\usepackage{multirow}
\usepackage{graphicx}
\usepackage{amsmath}
\usepackage{color}
\usepackage{mathrsfs}
\usepackage{float}
\usepackage{indentfirst}
\usepackage{mathrsfs}
\usepackage{float}
\usepackage{indentfirst}
\usepackage{textcomp}
\usepackage{comment}
\usepackage{mathtools}
\usepackage{natbib,hyperref}
\usepackage{soul}

\begin{document}
\title{Clifford circuits Augmented Matrix Product States for fermion systems}
%%%%%%%%%%%%

%%%%%%%%%%%%
\author{Jiale Huang}
\altaffiliation{These authors contributed equally to this work.}
\affiliation{Key Laboratory of Artificial Structures and Quantum Control (Ministry of Education),  School of Physics and Astronomy, Shanghai Jiao Tong University, Shanghai 200240, China}

\author{Xiangjian Qian}
\altaffiliation{These authors contributed equally to this work.}
\affiliation{Key Laboratory of Artificial Structures and Quantum Control (Ministry of Education),  School of Physics and Astronomy, Shanghai Jiao Tong University, Shanghai 200240, China}

\author{Mingpu Qin} \thanks{qinmingpu@sjtu.edu.cn}
\affiliation{Key Laboratory of Artificial Structures and Quantum Control (Ministry of Education),  School of Physics and Astronomy, Shanghai Jiao Tong University, Shanghai 200240, China}

\affiliation{Hefei National Laboratory, Hefei 230088, China}

\date{\today}

%%%%%%%%%%%%

%%%%%%%%%%%%
\begin{abstract}
    Clifford circuits Augmented Matrix Product States (CAMPS) was recently proposed to leverage the advantages of both Clifford circuits and Matrix Product States (MPS). Clifford circuits can support large entanglement and can be efficiently simulated classically according to the Gottesman-Knill theorem. So in CAMPS, MPS needs only to handle the so-called Non-stabilizerness Entanglement Entropy which significantly improves the simulation accuracy for a given bond dimension. In this work, we generalize CAMPS to study the Fermion system by taking advantage of the Jordan-Wigner transformation which can map the studied Fermion system to a spin system. We benchmark the method on both the spinless $t-V$ model and the spinful Hubbard model. Our test results show significant improvement of the accuracy of CAMPS over MPS, especially when the interactions are strong. Fermionic CAMPS provides a useful tool for the accurate study of many-body fermion systems in the future and has the potential to help resolve long-standing issues.
\end{abstract}
%%%%%%%%%%%%

%%%%%%%%%%%%
\maketitle
{\em Introduction --}
The study of strongly correlated many-body fermion systems is a cornerstone of modern condensed matter physics, underpinning our understanding of phenomena such as superconductivity, magnetism, and quantum phase transitions \cite{RevModPhys.66.763,qin2022hubbard,annurev-conmatphys-031620-102024}. However, simulating these systems presents significant challenges due to the complexity of fermion statistics and the exponential scaling of the Hilbert space with system size. Nowadays, the studies of these system mainly rely on numerical many-body method and numerous numerical methods have been developed over decades to tackle this challenge, such as quantum Monte Carlo methods \cite{RevModPhys.87.1067,RevModPhys.73.33}, Embedding methods (e.g., Dynamical Mean Field Theory \cite{RevModPhys.68.13, RevModPhys.78.865}, Density Matrix Embedding Theory \cite{PhysRevLett.109.186404}), tensor network methods \cite{PhysRevLett.69.2863,RevModPhys.93.045003,xiang2023density} and so on \cite{PhysRevX.5.041041}. However, each method has its limitation, necessitating the development of more efficient computational approaches to push the boundaries of our understanding of strongly correlated many-body fermion systems \cite{PhysRevX.5.041041}.

Density Matrix Renormalization Group (DMRG) or Matrix Product States (MPS) have emerged as a powerful tool for simulating (quasi) one-dimensional quantum many-body systems \cite{PhysRevLett.69.2863, SCHOLLWOCK201196}. However, when applied to high-dimensional systems, the restriction on the underlying entanglement structure of the ansatz makes the computational cost prohibitive. People also proposed two-dimensional generalization of MPS which can encode the entanglement structure of two-dimensional (2D) systems, such as Projected Entangled Pair States (PEPS) \cite{2004cond.mat..7066V,RevModPhys.93.045003,ORUS2014117}, 2D Multiscale Entanglement Renormalization Ansatz (MERA) \cite{PhysRevLett.102.180406,PhysRevLett.99.220405}, Projected Entangled Simplex States (PESS) \cite{PhysRevX.4.011025}, and so on \cite{annurev-conmatphys-040721-022705,xiang2023density,PhysRevLett.97.107206}. However, the cost of these methods is usually very high, preventing the reach of large bond dimensions. To address the limitation on entanglement and maintain the low cost of 
MPS, we propose the Clifford circuits Augmented Matrix Product States (CAMPS) method \cite{PhysRevLett.133.190402}, which seamlessly integrates the advantages of both Clifford circuits and MPS. This method arises from the Fully-augmented Matrix Product States (FAMPS) ansatz proposed in~\cite{Qian_2023}, where MPS are augmented with disentanglers to increase the encoded entanglement. In CAMPS, we choose the disentanglers as Clifford circuits instead of the general unitary circuits (as in FAMPS) to make the augmentation of MPS with multiple layers of disentanglers feasible. According to the Gottesman-Knill theorem \cite{gottesman1997stabilizer,PhysRevA.70.052328,PhysRevA.73.022334}, Clifford circuits can be simulated efficiently on classical computers, allowing them to handle the related entanglement entropy effectively \cite{PhysRevX.7.031016}. In CAMPS, this capability enables MPS to focus on the so-called Non-stabilizerness Entanglement Entropy (NsEE) \cite{nsee}, significantly improving simulation accuracy over MPS with a given bond dimension.

Initially, the CAMPS method was developed for ground-state calculations of spin systems, and it has been shown to significantly reduce the entanglement entropy and improve the simulation accuracy \cite{PhysRevLett.133.190402}. Subsequently, by incorporating Clifford circuits into the Time-Dependent Variational Principle (TDVP) framework \cite{PhysRevLett.107.070601,PhysRevB.94.165116}, the CAMPS method was generalized to time evolution \cite{qian2024cliffordcircuitsaugmentedtimedependent,Antonio0701} and finite-temperature simulations \cite{qian2024augmentingfinitetemperaturetensor}.  Moreover, the CAMPS method can be utilized to disentangle critical quantum chains \cite{fan2024disentanglingcriticalquantumspin,2024arXiv241111720F} and to provide new insights into the hardness of simulating quantum states classically \cite{nsee}. These works demonstrate that the improvements achieved by the CAMPS method are robust for different applications and highlight its potential for the accurate simulation of quantum many-body systems.

Previous works have focused exclusively on spin systems because Clifford circuits act on the Pauli basis. It is natural to ask whether the CAMPS method can also be applied to fermionic systems and whether similar advantages can be obtained. Understanding fermionic systems, such as the Hubbard model, is crucial for uncovering the physics of high-temperature superconductors and other exotic materials \cite{qin2022hubbard,annurev-conmatphys-031620-102024,PhysRevX.10.031016,doi:10.1126/science.aam7127,doi:10.1126/science.adh7691,PhysRevResearch.2.033073}. In this work, we present an initial exploration of the method on fermion systems. By taking advantage of Jordan-Wigner transformation \cite{jordan1928paulische}, we can map fermionic operators to spin operators, enabling us to apply the CAMPS method to fermionic systems. We test the method on both the spinless t-V model and the spinful Hubbard model. The benchmark results demonstrate that CAMPS can significantly improve computational efficiency and reduce entanglement entropy in fermionic systems compared to the standard MPS method. Our findings provide a promising avenue for future research on fermionic systems by leveraging the advantages of the CAMPS method.

{\em CAMPS for fermion systems--}
Since the ordinary Clifford circuits are defined in the spin $1/2$ basis, we need to map fermionic operators to spin operators to study fermion systems. The Jordan-Wigner transformation \cite{jordan1928paulische} is a common choice to perform the mapping. We notice that in addition to the Jordan-Wigner transformation, there are several other schemes available for mapping fermionic operators to spin operators. One prominent alternative is the Bravyi-Kitaev transformation \cite{Bravyi_2002}, which offers unique advantages in certain contexts. Recent studies \cite{PRXQuantum.5.030333,PhysRevA.98.032309,PhysRevResearch.5.023174,PhysRevB.92.075132,PhysRevC.108.024313} also explore various other techniques to improve the efficiency of the fermion-to-qubit mappings. 
Here we choose the Jordan-Wigner transformation for simplicity.

\begin{figure}[t]
    \includegraphics[width=80mm]{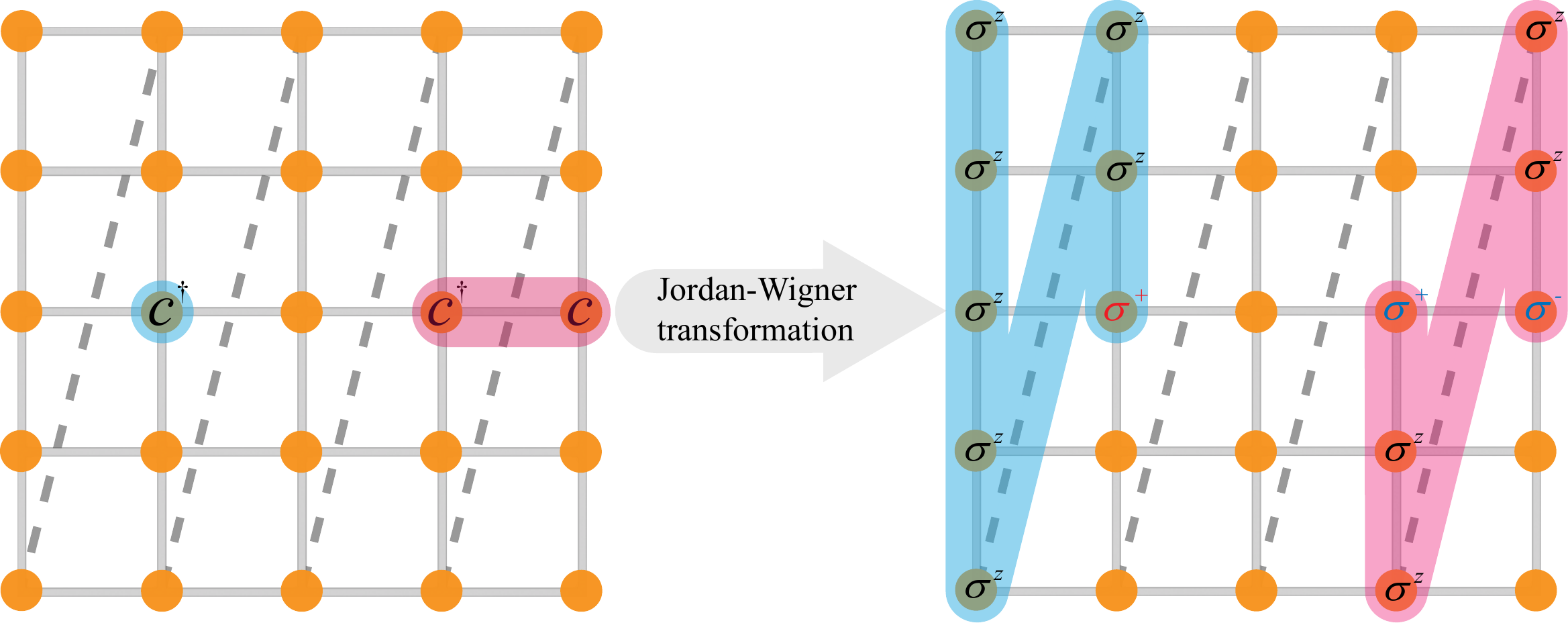}
       \caption{An illustration of the Jordan-Wigner transformation for a spinless fermion system on a two-dimensional lattice. We first employ the snake-like mapping to transform the 2D lattice into a 1D chain as denoted by the gray dashed line. The fermion creation operator $c^{\dagger}$ act on site $i$ is mapped to a product of Pauli operators: a string of  $\sigma^z$ operators from site $1$ to site $i-1$, followed by $\sigma^+$ at site $i$ which is highlighted in blue shadow. The hopping term between sites $i$ and $j$, highlighted in red shadow, is mapped to a product of Pauli operators consisting of $\sigma^+$ at site $i$, $\sigma^-$ at site $j$ and $\sigma^z$ operators inserted at all intermediate sites along the chain between $i$ and $j$. } 
       \label{JW_Trans}
\end{figure}

The fermionic creation $c^{\dagger}_i$ (annihilation $c_i$) operator creates (annihilates) a fermion at site $i$. They satisfies the anti-commutation relation:
\begin{equation}
    \{c_i,c_j\}=\{c^{\dagger}_i,c^{\dagger}_j\}=0, \{c_i,c^{\dagger}_j\}=\delta_{ij}
\end{equation}
The Jordan-Wigner transformation maps those operators to string-like operators acting on the spin space: 
\begin{equation}
    \begin{split}
    c^{\dagger}_i=& \sigma^z_1\sigma^z_2\cdots \sigma^z_{i-1}\sigma^+_i \\ 
    c_i =& \sigma^z_1\sigma^z_2\cdots \sigma^z_{i-1}\sigma^-_i 
    \end{split}
    \label{JW}
\end{equation}
where $\sigma^{\pm}_i=(\sigma^x_i\pm i\sigma^y_i)/2$, and $\sigma^{\alpha}_i$ are the Pauli matrices acting on site $i$. Accordingly, the density operator $n_i=c^{\dagger}_ic_i$ is mapped to $(1+\sigma^z_i)/2$. The transformation in Eq.~(\ref{JW}) is defined for spinless fermions, but it can be easily generalized to spinful fermions (for example, in the Hubbard model) by treating the electron operators with different spin index as different electron operators. We notice that for one-dimensional fermion system, the long-range string operator in Eq.~(\ref{JW}) is usually canceled in the resulting spin Hamiltonian, but it remains in higher-dimension cases. Fig.~\ref{JW_Trans} illustrates the Jordan-Wigner transformation applied to a spinless fermion system on a two-dimensional lattice. To perform the Jordan-Wigner transformation on a 2D lattice, it is necessary to map the lattice into a one-dimensional chain first. In this work, we employ the common snake-like mapping as shown in Fig.~\ref{JW_Trans}. We do not provide the explicit transformation for the interaction terms here as they can be readily obtained by applying the transformations outlined in Eq.~(\ref{JW}).
With the Jordan-Wigner transformation, we can map any fermionic Hamiltonian to a spin Hamiltonian as: 
\begin{equation}
    H = \sum_{i} a_i P_i
    \label{H}
\end{equation}
where $a_i$ are the coefficients, $P_i$ are the Pauli strings $P=\sigma_1 \otimes \sigma_2\cdots \otimes \sigma_N$ ($\sigma_i\in \{I, \sigma^x, \sigma^y, \sigma^z\}$). Then we can easily apply the CAMPS method to this spin Hamiltonian. Since U(1) symmetry is not imposed in the Clifford circuits, a chemical potential term is included, which is tuned to target the desired filling factor.

\begin{table}
    \centering
    \begin{tabular}{cccccc} 
    \hline
    model & $L_x\times L_y$ &  $V$ ($U$) & doping &  $D$ & $E_\text{ref}$ \\
    \hline
    \multirow{6}{*}{$t-V$} & $6\times 6$ & $0.5$ & $0$ & $10000$ & $- 0.7900534$ \\
    & $6\times 6$ & $1$ & $0$ & $10000$ & $-0.8735384$ \\
    & $6\times 6$ & $3$ & $0$ & $4000$ & $-1.467808$ \\
    \cline{2-6}
    & $8\times 8$ & $0.5$ & $0$ & $20000$ & $-0.8101752$ \\
    & $8\times 8$ & $1$ & $0$ & $20000$ & $-0.8961414$ \\
    & $8\times 8$ & $3$ & $0$ & $8000$ & $-1.527751$ \\
    \hline
    \multirow{3}{*}{Hubbard} & $1\times 32$ & 8 & 1/8 & $600$ & $-1.738557$ \\
    & $2\times 16$ & 8 & 1/8 & $6000$ & $-1.977745$ \\
    & $4\times 8$ & 8 & 1/8 & $40000$ & $-2.070021$ \\
    \hline
    \end{tabular}
    \caption{The reference ground state energy for the 2D $t-V$ model at half-filling and the Hubbard model at 1/8 doping. The energy unit is set to $t=1$. All results are obtained by DMRG with open boundary conditions. The listed significant digits of energies are checked by the extrapolation of energies with truncation errors using large bond dimension values.}
    \label{energy_reference}
\end{table}

The MPS ansatz is defined as:
\begin{equation}
    |\text{MPS}\rangle = \sum_{\{\sigma_i\}} \text{Tr}(A^{\sigma_1}_1A^{\sigma_2}_2\cdots A^{\sigma_N}_N)|\sigma_1\sigma_2\cdots \sigma_N\rangle
    \label{MPS}
\end{equation}
where $A$ is a rank-3 tensor with a physical index $\sigma_i$ (with dimension $2$ for spin $1/2$ degree of freedom) and two auxiliary indices with dimension $D$. By applying Clifford circuits to the MPS wave function, we obtain a new wave function, termed Clifford circuits augmented Matrix Product States  (CAMPS) \cite{PhysRevLett.133.190402,2024arXiv240418751L}, as $|\text{CAMPS}\rangle = \mathcal{C}|\text{MPS}\rangle$, where $\mathcal{C}$ denotes the Clifford circuits. 

One advantage of the CAMPS method is its ability to efficiently compute the expectation value of any observable of interest. 
%ensuring that physical quantities remain invariant under the action of the Clifford circuit $\mathcal{C}$. 
The expectation value of any observable $O=\sum_i P_i$ ($P_i$ represents a specific type of Pauli string) can be calculated as $\langle O\rangle=\langle\text{CAMPS}|O|\text{CAMPS}\rangle=\langle\text{MPS}|O'|\text{MPS}\rangle$, where $O' = \mathcal{C}^\dagger O \mathcal{C}$. Since Clifford circuits transform a Pauli string into another Pauli string, $O^\prime$ maintains the same level of simplicity as $O$, involving an equivalent number of summations of Pauli strings, ensuring that the calculation of the CAMPS remains quite efficient. In contrast to Clifford circuits, applying general unitary circuits to MPS causes the summation terms in $O^\prime$ to increase exponentially with the number of layers of the unitary circuits, rendering the computation infeasible. In CAMPS \cite{PhysRevLett.133.190402}, the local MPS matrices, the local Clifford circuits and the ansatz structure itself can be optimized simultaneously by slightly modifying the usual two-site DMRG algorithm. More details can be found in \cite{PhysRevLett.133.190402}.

\begin{figure}[t]
    \includegraphics[width=80mm]{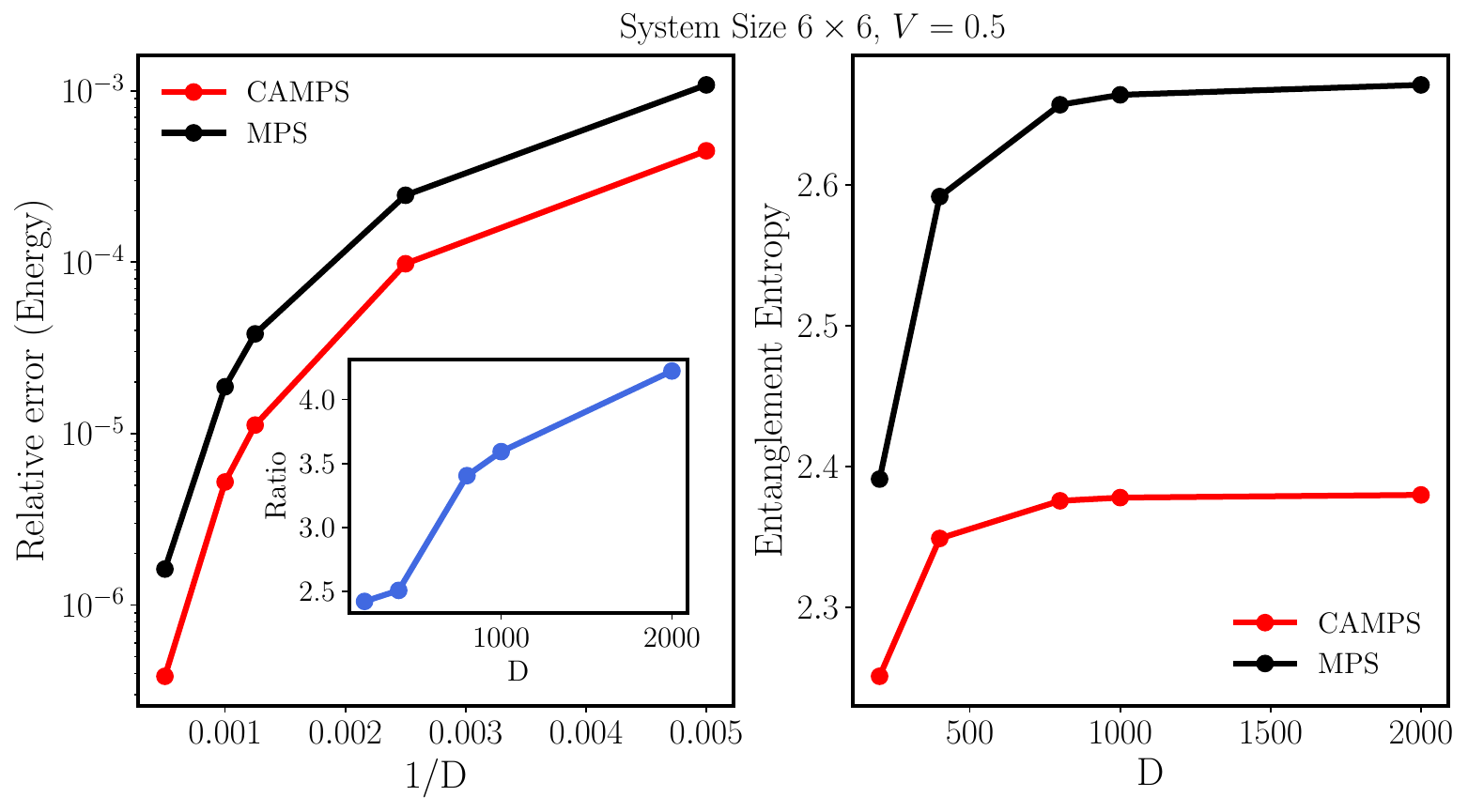}
    \includegraphics[width=80mm]{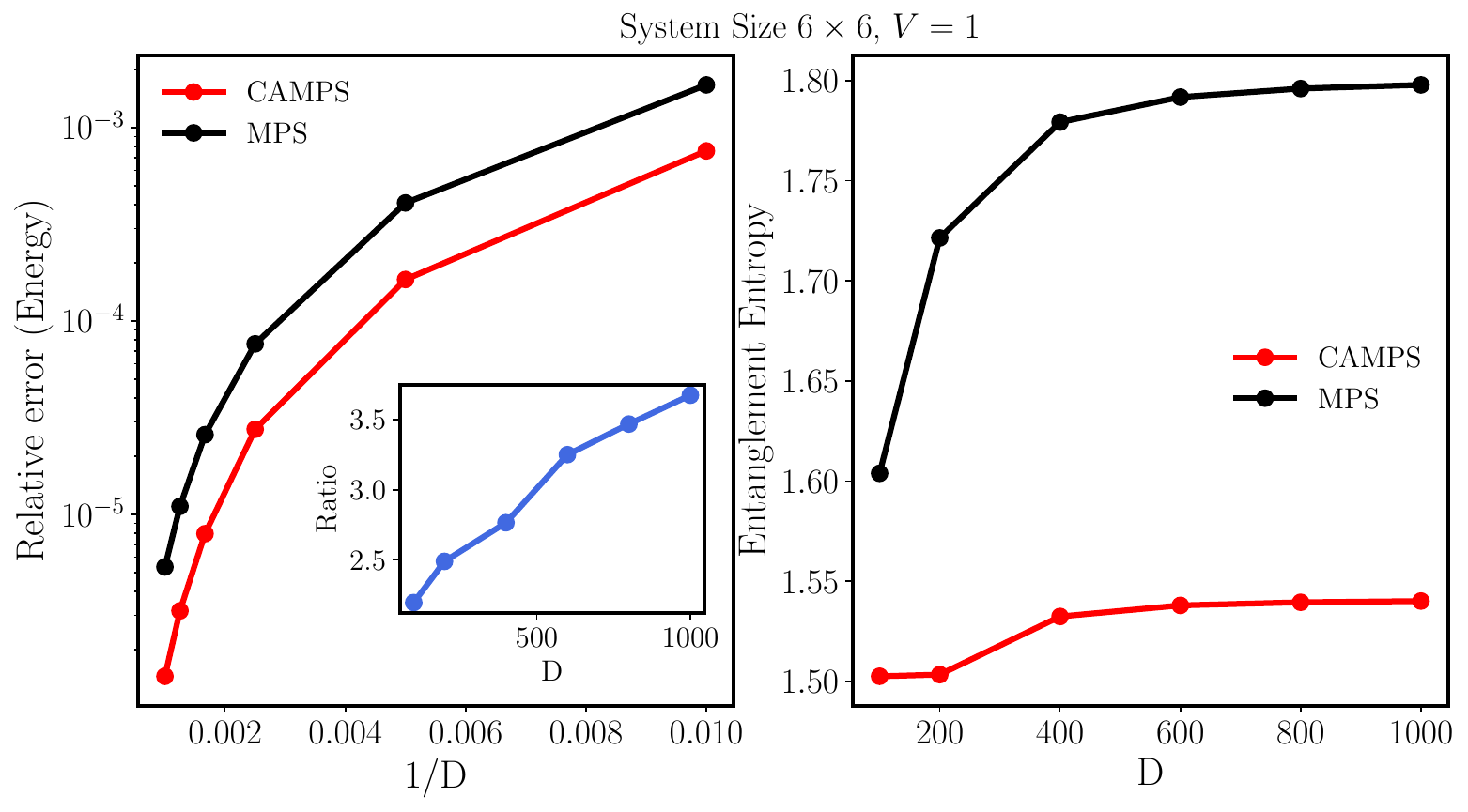}
    \includegraphics[width=80mm]{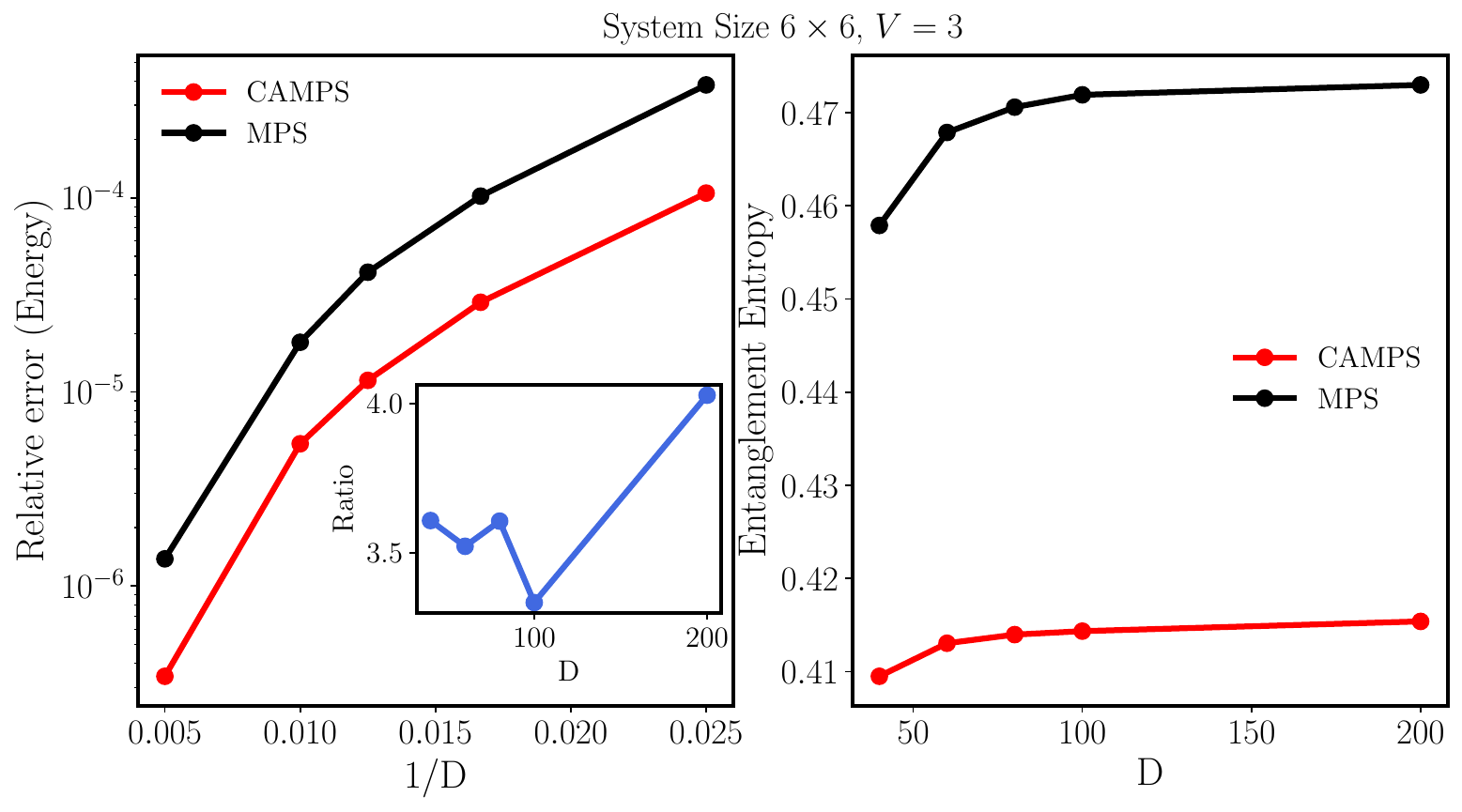}
       \caption{Results of 2D $t-V$ model at half filling with size $6\times 6$.
        Relative error of the ground state energy and the Entanglement entropy at the center bond in the MPS part for CAMPS and MPS as a function of bond-dimension $D $ for different $V$ are shown. The reference ground state energy can be found in Table~\ref{energy_reference}.
        The ratio of relative error $\frac{E_\text{MPS} - E_\text{ref}}{E_\text{CAMPS} - E_\text{ref}}$ of the ground state energy between MPS and CAMPS is shown in the inset of the left panels. } 
       \label{TV_6}
\end{figure}

\begin{figure}[t]
    \includegraphics[width=80mm]{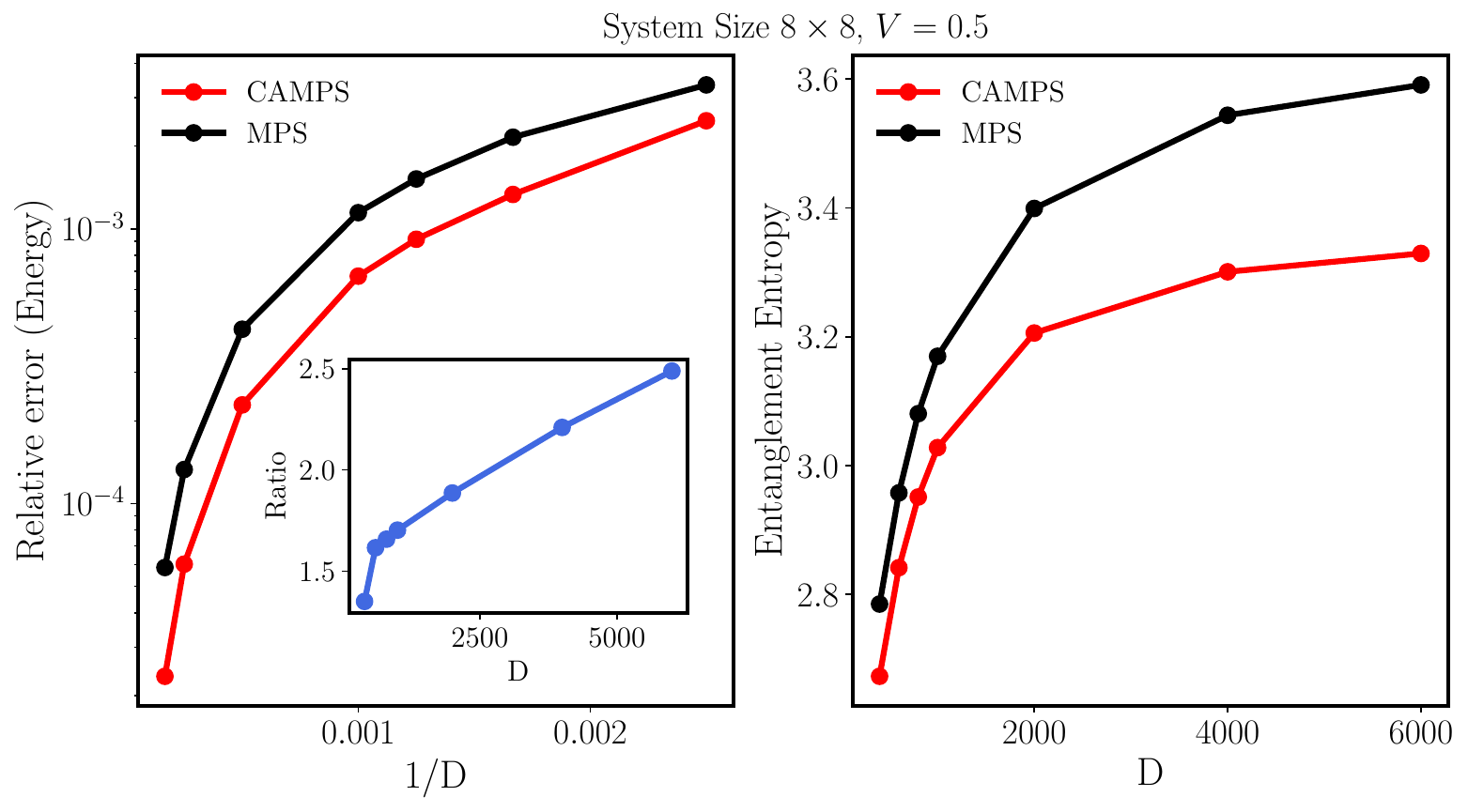}
    \includegraphics[width=80mm]{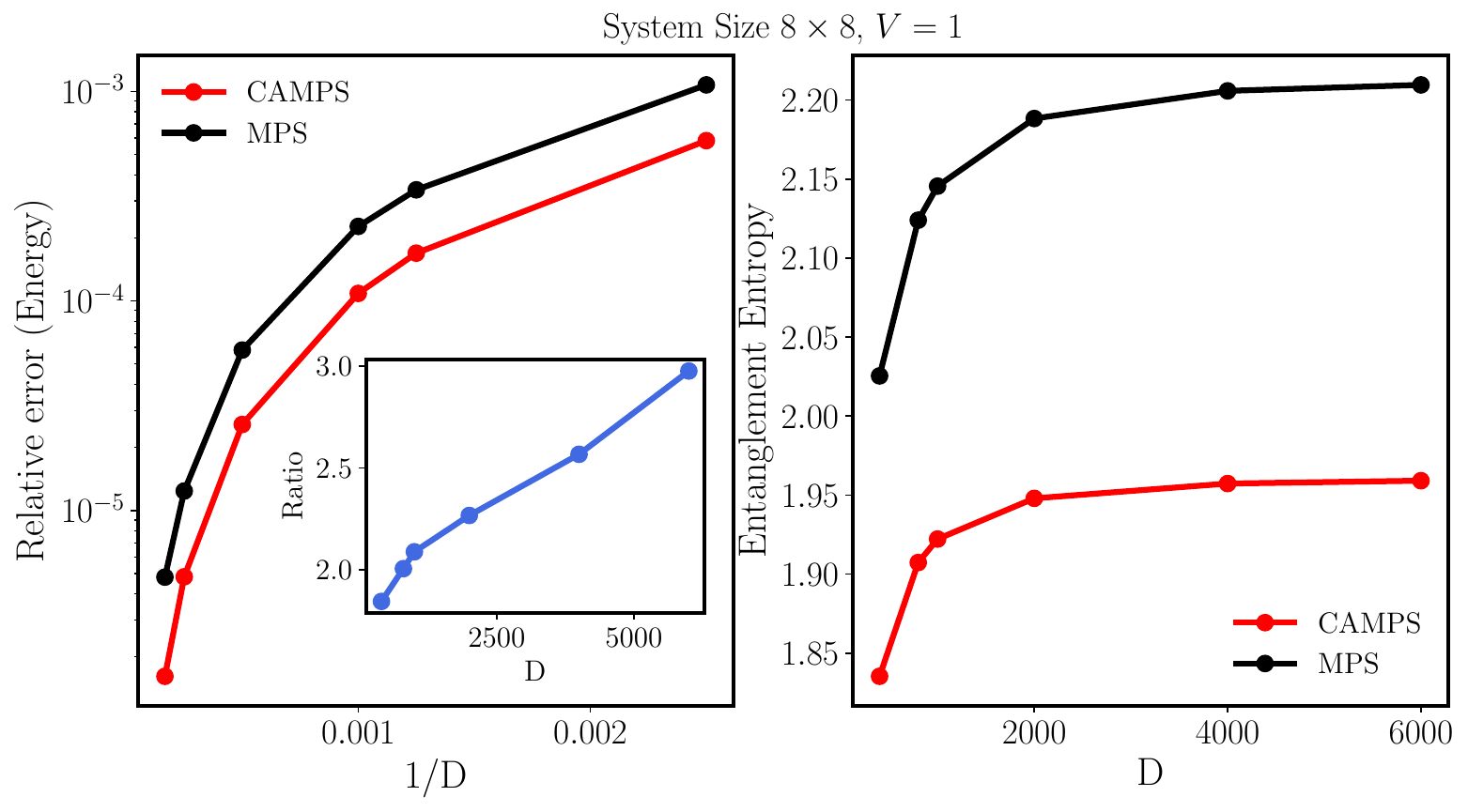}
    \includegraphics[width=80mm]{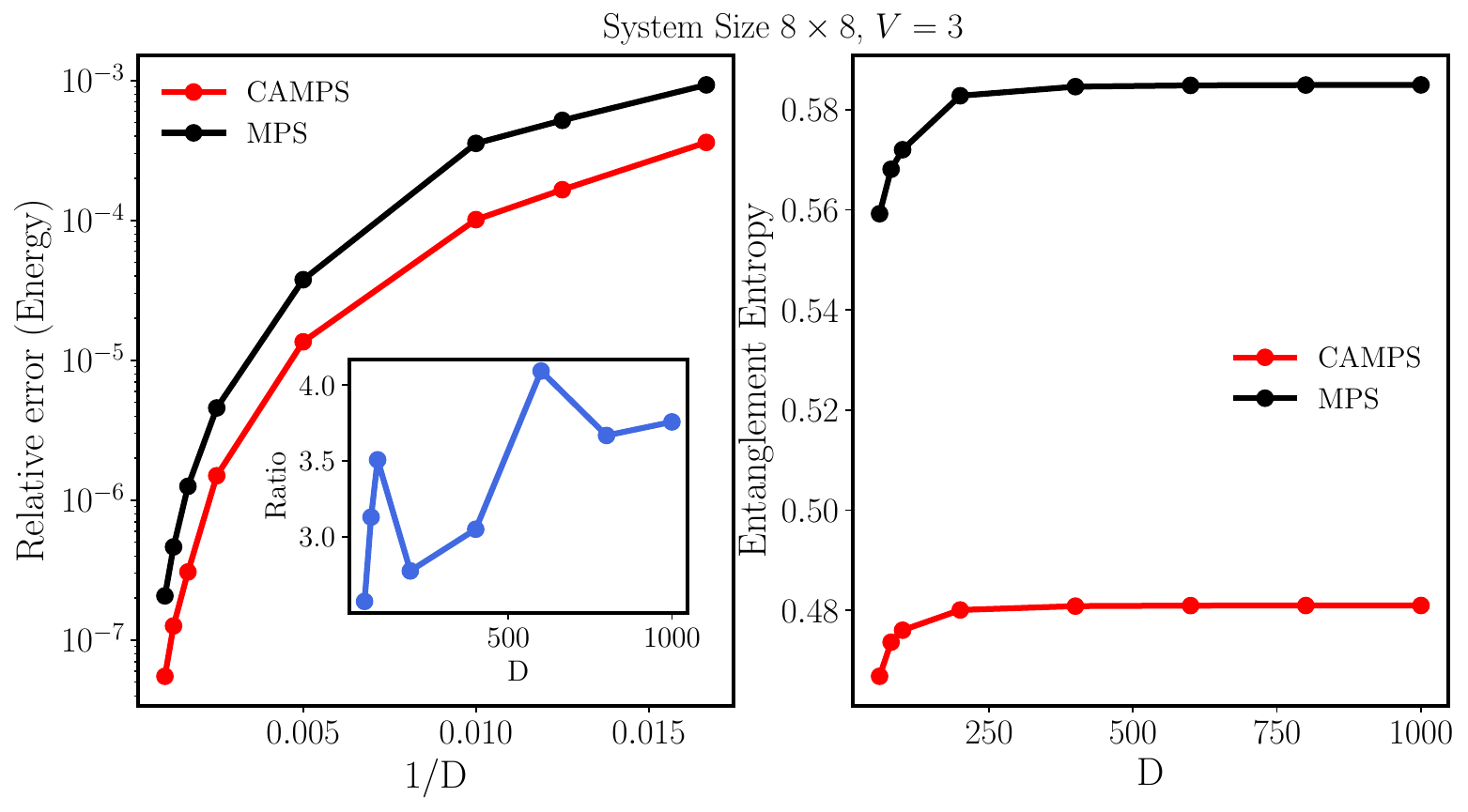}
       \caption{Similar as Fig.~\ref{TV_6}. Results for 2D $t-V$ model at half filling with size $8\times 8$.
        Relative error of the ground state energy and the Entanglement entropy at the center bond in the MPS part for CAMPS and MPS as a function of bond-dimension $D $ for different V are shown. The reference ground state energy can be found in Table~\ref{energy_reference}. 
        The ratio of Relative error $\frac{E_\text{MPS} - E_\text{ref}}{E_\text{CAMPS} - E_\text{ref}}$ of the ground state energy between MPS and CAMPS is shown in the inset of the left panels.} 
       \label{TV_8}
\end{figure}

\begin{figure}[t]
    \includegraphics[width=80mm]{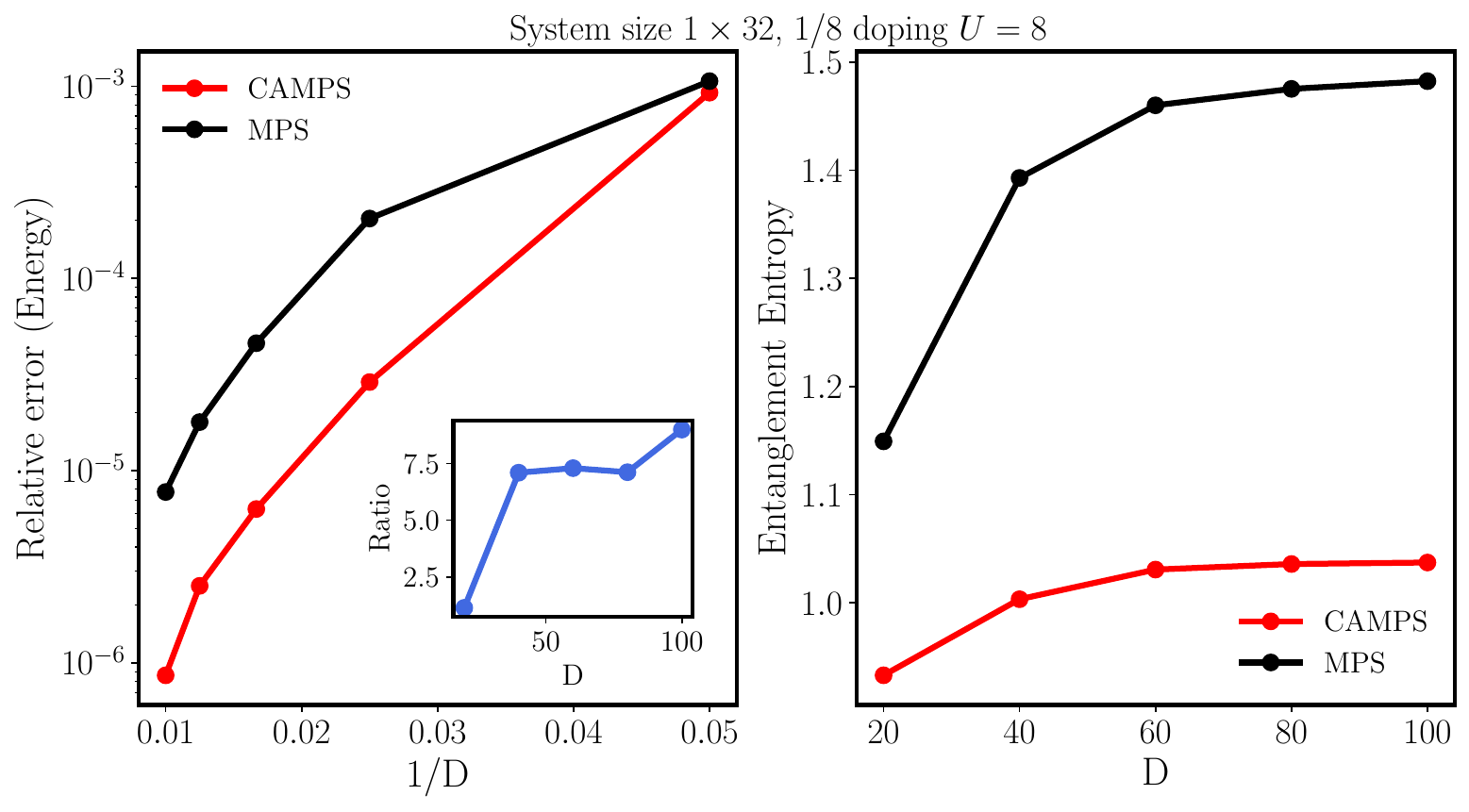}
    \includegraphics[width=80mm]{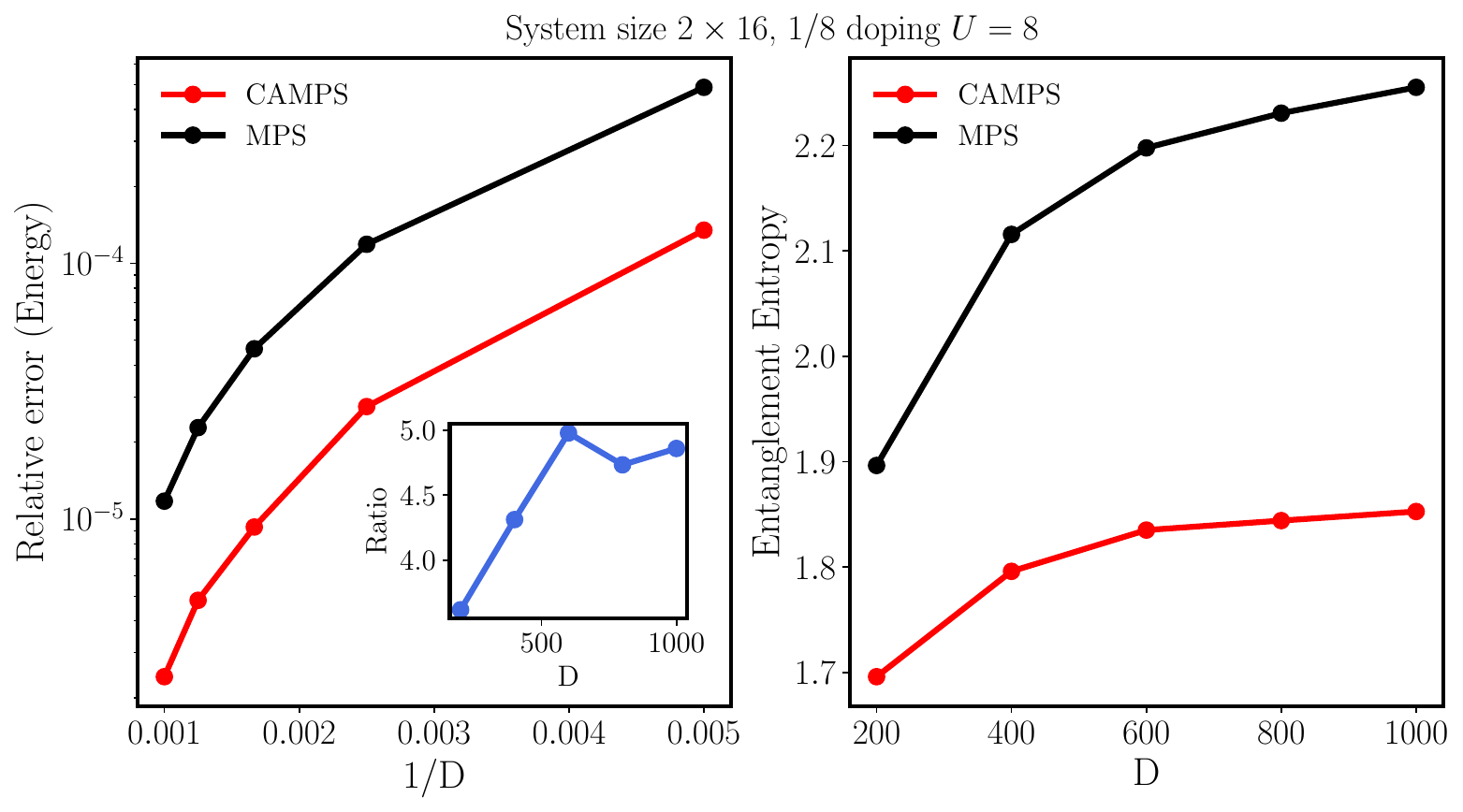}
    \includegraphics[width=80mm]{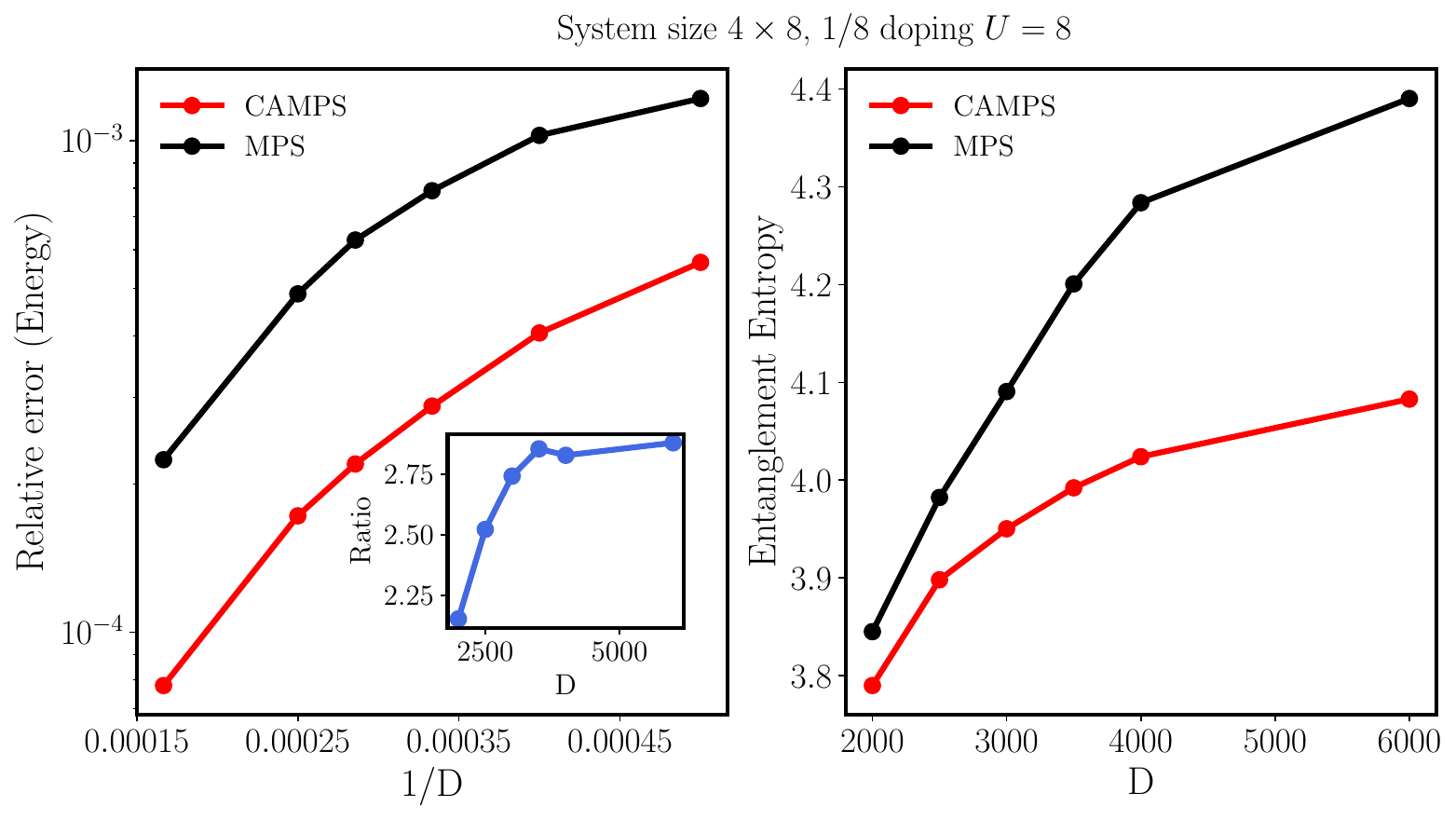}
       \caption{ Results for the Hubbard model at $1/8$ hole doping with $U = 8$.
       Relative error of the ground state energy and the entanglement entropy at the center bond in the MPS part for CAMPS and MPS as a function of bond-dimension $D$ for different lattice sizes are shown. The reference ground state energy can be found in Table~\ref{energy_reference}. The ratio of Relative error $\frac{E_\text{MPS} - E_\text{ref}}{E_\text{CAMPS} - E_\text{ref}}$ of the ground state energy between MPS and CAMPS is shown in the inset of the left panels. }
       \label{Hubbard}
\end{figure}

{\em Results for 2D $t-V$ model --}
The $t-V$ model is a spinless fermion model with Hamiltonian
\begin{equation}
	H = -t\sum_{\langle i, j\rangle} c^\dagger_ic_j + h.c. + V\sum_{\langle i, j\rangle}(n_i - 1/2) (n_j-1/2)
 \label{TV_2D_H}
\end{equation}
where $c_i$ ($c_i^\dagger$) is the annihilation (creation) operator of a spinless fermion at site $i$, $n_i=c_i^\dagger c_i$ is the density operator, and $t$ and $V$ are the hopping and interaction parameters, respectively. The sum $\langle i, j\rangle$ runs over nearest-neighbor sites. We set $t=1$ as the energy unit in our calculation. 

The particle-hole symmetry of this model makes the Hamiltonian defined in Eq.~(\ref{TV_2D_H}) half-filling, where the ground state shows check-board charge density wave \cite{PhysRevB.81.165104,PhysRevB.102.205122}. The model displays phase separation slightly away from half-filling \cite{PhysRevB.81.165104,PhysRevB.102.205122}. We focus on the half-filling case in this work. We consider two system sizes $6\times 6$ and $8\times 8$. Open boundary conditions are imposed for simplicity. For both system sizes, we consider three different values of the interaction strength $V=0.5$, $1$, and $3$. We utilize the energies obtained by DMRG with very large bond dimensions as the reference which can be found in Table~\ref{energy_reference}. The listed significant digits of energies in Table~\ref{energy_reference} are checked by the extrapolation of energies with truncation errors using large bond dimension values.

The relative errors of the ground-state energy and the entanglement entropy at the central bond in the MPS part, for both CAMPS (i.e., NsEE \cite{nsee}) and MPS, are shown in Figs.~\ref{TV_6} and \ref{TV_8}. For both system sizes, the relative errors are reduced significantly with CAMPS, and the reduction becomes more dramatic with the increase of $V$. 
For the system with size $8\times 8$ and $V=3$, the relative error can be reduced by a ratio factor (defined as $\frac{E_\text{MPS} - E_\text{ref}}{E_\text{CAMPS} - E_\text{ref}}$) of $4$, and this factor continues to increase with bond dimension $D$. The entanglement entropy is also reduced in the MPS part of CAMPS calculations compared to pure MPS calculations, as shown in Fig.~\ref{TV_6} and Fig.~\ref{TV_8}.

{\em Results for the Hubbard model --}
We also test the CAMPS method on the Hubbard model, with Hamiltonian
\begin{equation}
	H = -t\sum_\sigma\sum_{\langle i, j\rangle}c^\dagger_{i,\sigma}c_{j,\sigma} + h.c. \\ \nonumber
    + U\sum_{i}n_{i,\uparrow}n_{i,\downarrow} - \mu\sum_{i}n_i
    \label{H_Hubbard}
\end{equation}
where $c_{i,\sigma}$ ($c_{i,\sigma}^\dagger$) is the annihilation (creation) operator of a spinful fermion at site $i$ with spin $\sigma$, $n_{i,\sigma}=c_{i,\sigma}^\dagger c_{i,\sigma}$ is the corresponding density operator, $n_i = n_{i,\uparrow} + n_{i,\downarrow}$, $t$ is the hopping parameter, $U$ is the on-site interaction, and $\mu$ is the chemical potential. We set $t=1$ as the energy unit, and the interaction strength is $U=8$ in our calculation.

We tune the chemical potential $\mu$ to set the system at 1/8 hole doping. We test the CAMPS method on the Hubbard model with lattice sizes $1 \times 32$, $2 \times 16$, and $4 \times 8$ under open boundary conditions. As mentioned earlier, after applying the Jordan-Wigner transformation, the system size is doubled in the resulting spin model. The simulation results are shown in Fig.~\ref{Hubbard}. We utilize the results obtained with large bond dimensions as the reference (can be found in Table~\ref{energy_reference}).
Similar to the $t-V$ model, the improvement in the relative error of the ground-state energy is significant, and the ratio factor $\frac{E_\text{MPS} - E_\text{ref}}{E_\text{CAMPS} - E_\text{ref}}$ increases with bond dimension $D$. We also observe a reduction in entanglement entropy across all system sizes for CAMPS calculations as shown in the inset of Fig.~\ref{Hubbard}.

{\em Conclusion and Perspective --}
In this work, we generalize the CAMPS method to fermion systems by mapping the fermion system to a spin one through the Jordan-Wigner transformation. Our test results on the $t-V$ and Hubbard model show that CAMPS can significantly improve accuracy compared to MPS. Moreover, the improvement becomes more dramatic with the increase of the bond dimension. The framework can be easily generalized to the real time evolution \cite{qian2024cliffordcircuitsaugmentedtimedependent,Antonio0701} and the finite temperature \cite{qian2024augmentingfinitetemperaturetensor} calcualtions. It will be interesting to test other schemes to map the fermionic degree to spin degree and see how the performance of CAMPS changes. All the calculations in this work are in the grand-canonical ensemble. It is necessary to impose the U(1) symmetry in the method to improve the simulation efficiency for system with particle number conservation. The fermionic CAMPS provides a useful approach for the accurate study of many-body fermion systems in the future. 

%%%%%%%%%%%%

\begin{acknowledgments}
The calculation in this work is carried out with TensorKit \cite{foot7}. The computation in this paper were run on the Siyuan-1 cluster supported by the Center for High Performance Computing at Shanghai Jiao Tong University. MQ acknowledges the support from the National Key Research and Development
Program of MOST of China (2022YFA1405400), the National Natural Science Foundation of China (Grant No. 12274290), the Innovation Program for Quantum Science and Technology (2021ZD0301902), and the sponsorship from Yangyang Development Fund.
%%%%%%%%%%%%
\end{acknowledgments}

%%%%%%%%%%%%Refs
\bibliography{main}
%%%%%%%%%%%%

\end{document}